\documentclass[twocolumn]{aastex62}
\usepackage{hyperref}
\usepackage{ulem}
\usepackage{graphicx,times}
\usepackage{epsfig}
\usepackage{xcolor}
\usepackage{rotating}
\newcommand{\maxi}{MAXI~J1348-630}
\newcommand{\hxmt}{{\it Insight}-HXMT}
\newcommand{\swift}{{\it Swift}}

\shorttitle{\maxi{} 2019 outburst}
\shortauthors{Weng~et~al.}
\begin{document}


\title{Time-lag between disk and corona radiation leads to hysteresis effect observed in black hole X-ray binary \maxi{}}

\author[0000-0001-7595-1458]{Shan-Shan~Weng}
\affil{Department of Physics and Institute of Theoretical Physics, Nanjing Normal University, Nanjing 210023, China}
\email{wengss@njnu.edu.cn}

\author[0000-0002-4223-2198]{Zhen-Yi~Cai}
\affil{CAS Key Laboratory for Research in Galaxies and Cosmology, Department of Astronomy, University of Science and Technology of China, Hefei 230026, China}
\affil{School of Astronomy and Space Science, University of Science and Technology of China, Hefei 230026, China}

\author[0000-0001-5586-1017]{Shuang-Nan~Zhang}
\affil{Key Laboratory of Particle Astrophysics, Institute of High Energy Physics, Chinese Academy of Sciences, Beijing 100049, China}
\email{zhangsn@ihep.ac.cn}
\footnote{Corresponding authors: Shan-Shan Weng and Shuang-Nan Zhang}

\author{Wei~Zhang}
\affil{Key Laboratory of Particle Astrophysics, Institute of High Energy Physics, Chinese Academy of Sciences, Beijing 100049, China}

\author{Yu-Peng~Chen}
\affil{Key Laboratory of Particle Astrophysics, Institute of High Energy Physics, Chinese Academy of Sciences, Beijing 100049, China}

\author{Yue~Huang}
\affil{Key Laboratory of Particle Astrophysics, Institute of High Energy Physics, Chinese Academy of Sciences, Beijing 100049, China}

\author[0000-0002-2705-4338]{Lian~Tao}
\affil{Key Laboratory of Particle Astrophysics, Institute of High Energy Physics, Chinese Academy of Sciences, Beijing 100049, China}

\begin{abstract}
Accretion is an essential physical process in black hole X-ray binaries (BHXRBs) and active galactic nuclei. The properties of accretion flows and their radiation were originally considered to be uniquely determined by the mass accretion rate of the disk; however, the ``hysteresis effect'' observed during outbursts of nearly all BHXRBs seriously challenges this paradigm. The hysteresis effect is referred to that the hard-to-soft state transition in the fast-rise stage occurs at much higher luminosity than the soft-to-hard state transition in the slow-decay stage. That is, the same source can show different spectral/temporal properties at the same luminosity. Phenomenologically, this effect is also represented as the so-called ``q''-shaped hardness-intensity diagram, which has been proposed as a unified scene for BHXRBs. However, there is still a lack of quantitative theoretical interpretation and observational understanding of the ``q''-diagram. Here, we present a detailed time-lag analysis of a recently found BHXRB, \maxi{}, intensively monitored by \hxmt{} over a broad energy band (1--150 keV). We find the first observational evidence that the observed time-lag between radiations of the accretion disk and the corona leads naturally to the hysteresis effect and the ``q''-diagram. Moreover, complemented by the quasi-simultaneous \swift{} data, we achieve a panorama of the accretion flow: the hard X-ray outburst from the corona heats and subsequently induces the optical brightening in the outer disk with nearly no lag; thereafter, the enhanced accretion in the outer disk propagates inward, generating the delayed soft X-ray outburst at the viscous timescale of $\sim 8-12$ days.

\end{abstract}


\section{Introduction}
Black hole X-ray binaries (BHXRBs) are systems in which a black hole accretes material from the companion star via the Roche-lobe overflows. The infalling gas forms an accretion disk, where the angular momentum transfers outward and the matter diffuses toward the central compact object through viscous interaction \citep[for reviews, see][]{frank2002, done2007}. The gravitational energy turns into electromagnetic radiation, mostly in the X-ray band. BHXRBs generally stay at very faint quiescent state for most of their lifetime, but they may enter into outbursts with luminosity increased by several orders of magnitude within a few days, and then decay on a timescale of a few months \citep[e.g.,][]{remillard2006, corra2016}. Additionally, BHXRBs can be bright in the optical and radio bands arising from the outer accretion disk and jet, respectively \citep[e.g.,][]{van1994, gallo2003, russell2006, rykoff2007}.

The transient behavior of BHXRBs is generally considered as the result of the thermal-viscous instability propagating in the accretion disk \citep[e.g.,][]{chen1997, dubus2001, coriat2012}. The disk instability model (DIM) provides a basic framework;  meanwhile, the accretion disk truncation and X-ray irradiation should be involved to reproduce the global properties of BHXRB outbursts \citep[for reviews, see][]{lasota2001, hameury2020}. Irradiation plays a crucial role in BHXRBs, provides additional heating, affects the thermal equilibrium, and stabilizes the disk \citep[e.g.,][]{dubus2001}. In addition to the ``fast-rise slow-decay'' type light curve, the irradiation theory predicts a scaling relation between the optical luminosity, $L_{\rm O}$, and X-ray luminosity, $L_{\rm X}$, \citep[i.e., $L_{\rm O} \propto L_{\rm X}^{0.5}$;][]{van1994}, which has been confirmed in some BHXRBs \citep[e.g.,][]{king1998, russell2006, rykoff2007}. It is worth noting that, with some modifications, the irradiation scenario likely operates in active galactic nuclei (AGNs) as well (e.g., \citealt{krolik1991,cackett2007,edelson2015}; however, see \citealt{zhu2018,cai2018,cai2020}).

The properties of accretion flows and their radiation are generally considered to be a monotonic function of the mass accretion rate of the disk \citep{esin1997, yuan2014}. The terminologies of the ``high/soft'' and ``low/hard'' states are widely used in literature \citep[e.g.,][]{remillard2006, done2007}. However, this oversimplified assumption is challenged by the discovery of the ``hysteresis effect'' in nearly all BHXRBs; that is, apparently similar spectral state transitions take place at very different luminosities during an outburst cycle \citep[e.g.,][]{maccarone2003}. Therefore, it directly leads to the ``q''-shaped hardness-intensity diagram (HID), which has been proposed as a unified scheme for BHXRBs \citep{fender2004, belloni2010}. High signal-to-noise ratio data show that there is no strict one-to-one correlation between the spectral/temporal parameters and the X-ray luminosity \citep[e.g.,][]{remillard2006}. Therefore, instead of solely using the X-ray luminosity, currently the accretion states of BHXRBs are explicitly defined by their spectral and the temporal properties. Many different properties of the accretion flow have been suspected to be responsible for the hysteresis effect, e.g., the size of corona \citep{homan2001}, the mass of accretion disk \citep{yu2007}, the history of truncated disk radius \citep{zdziarski2004}, and the hydrogen ionization instability \citep{done2007}. However, none of these models can fully account for the hysteresis effect and the ``q''-shaped HID.

\maxi{} was first discovered by {\it MAXI} on 2019 January 26 due to its X-ray outburst. Soon afterward \hxmt{}, \swift{}, {\it NICER}, and {\it INTEGRAL} performed target-of-opportunity (ToO) observations to monitor the outburst and the subsequent re-flares \citep[e.g.,][]{chen2019, bassi2019, lepingwell2019, sanna2019}. Several groups had also carried out optical and radio observations \citep[e.g.,][]{carotenuto2019, charles2019, russell2019b, russell2019a}. The spectral/temporal properties of \maxi{}, including the ``q''-shaped HID \citep[e.g.,][]{tominaga2020}, the low-frequency quasi-periodic oscillations evolving with the spectral states \citep{jana2020, belloni2020, zhangliang2020}, and the radio flare occurred at the state transition \citep{carotenuto2019}, are all well consistent with those of the classical BHXRBs. Taking advantage of the broadband coverage of \hxmt{} and \swift{} on \maxi{}, we aim to investigate the mechanism for the hysteresis behavior and the scaling relations, and test the reprocessing scenario. The \hxmt{} and \swift{} data reduction is described in the next section, and the results are presented in Section 3. Discussion and conclusions follow in Section 4.

\section{Data reduction}
\maxi{} entered into an outburst in 2019 January, then decayed for $\sim 4$ months, and then re-brightened twice in the second half of 2019. In this work, we focus on the primary outburst, which was better sampled by both \hxmt{} and \swift{}.

\begin{figure*}
\centering
\includegraphics[width=\textwidth]{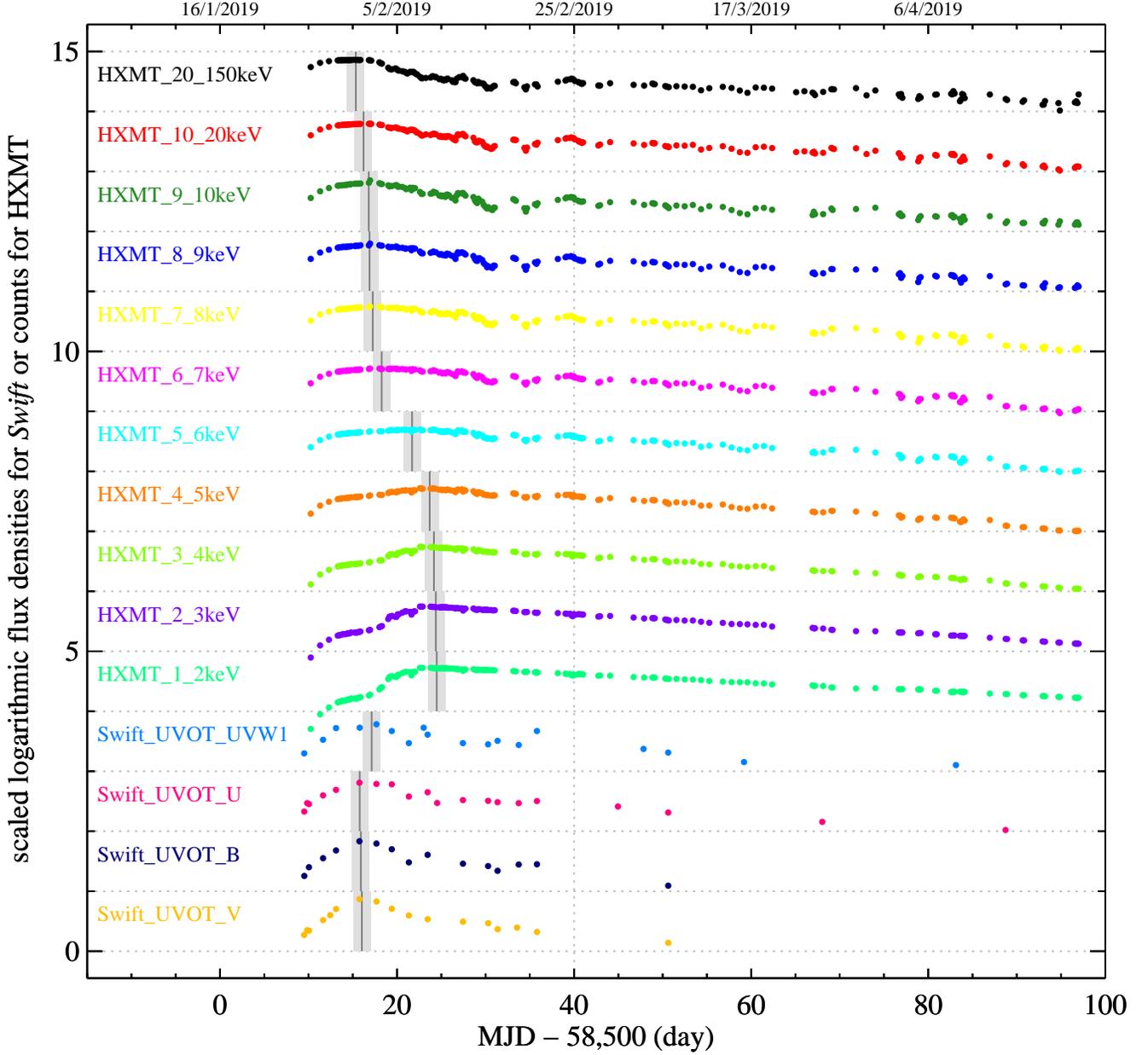}
\caption{Light curves of \maxi{}. The original {\it Insight}-HXMT fluxes are given in units of count s$^{-1}$ and the {\it Swift} flux densities in mJy, which are rescaled in logarithm such that their standard deviations are 0.2 and shifted vertically for clarity. The first prominent peak of each light curve determined by a fourth-order polynomial fit is marked as the gray solid vertical line with uncertainty of $\pm 1$ day in the light-gray region. The light curves before the light-gray dotted vertical line are adopted to estimate their lags relative to the one in the shortest wavelength band, i.e., the {\it Insight}-HXMT 20-150 keV band, using the linearly interpolated cross-correlation method. \label{fig:lc}}
\end{figure*}

\begin{figure}
\centering
\includegraphics[width=0.45\textwidth]{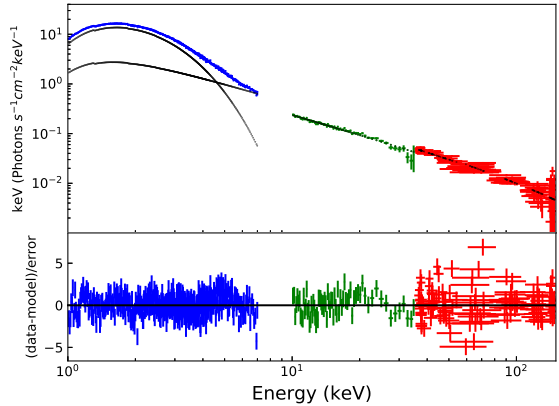}
\caption{Example spectrum of \maxi{}. Blue, green, and red symbols correspond to the LE, ME, and HE data, respectively. The disk and the cutoff power-law components are marked with dotted lines. \label{fig:spec}}
\end{figure}

\begin{figure*}
\centering
\includegraphics[width=0.9\textwidth]{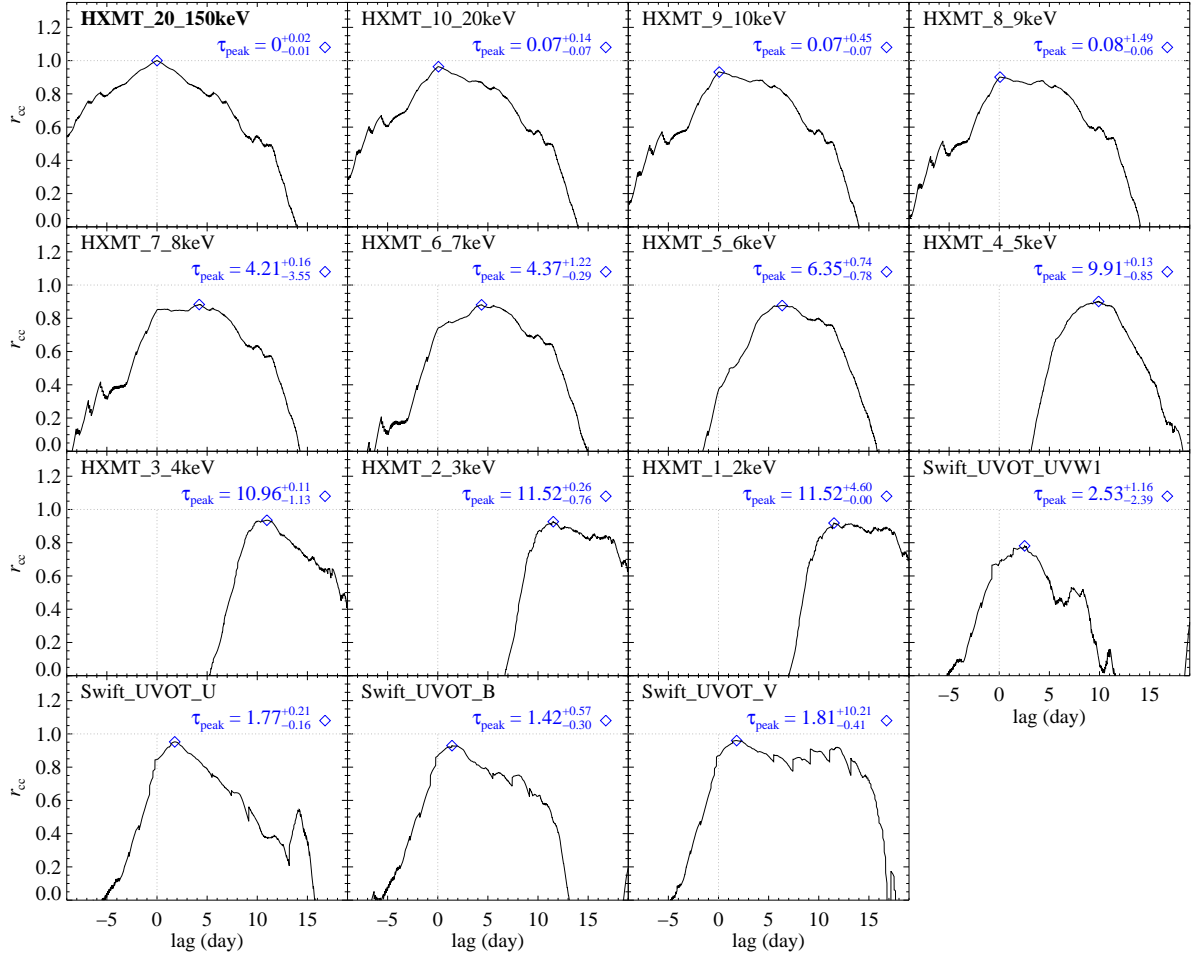}
\caption{Cross-correlation functions for the light curves relative to the one at the shortest wavelength, i.e., the HXMT\_20\_150keV in boldface. The peak lag, $\tau_{\rm peak}$, at the maximum of the cross-correlation function is nominated as the blue diamond symbol.}\label{fig:ccf}
\end{figure*}

\begin{figure}
\centering
\includegraphics[width=0.45\textwidth]{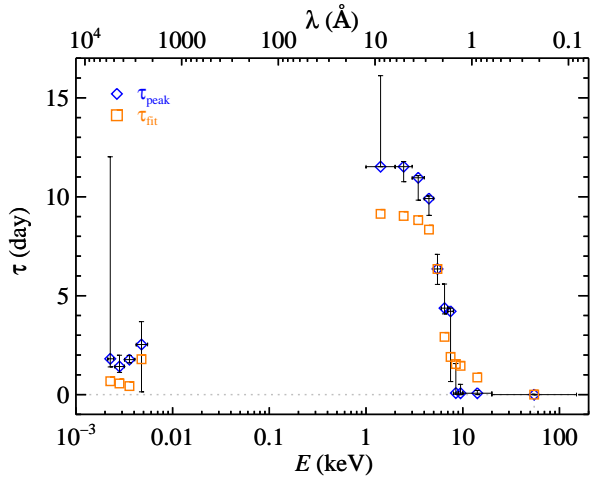}
\caption{
Lag-energy relation inferred from the multi-wavelength light curves relative to the HE light curve. $\tau_{\rm fit}$ (orange squares with a typical uncertainty of $\sim 1$ day) and $\tau_{\rm peak}$ (blue diamonds with 10-90th percentile uncertainties) are determined using the fourth-order polynomial fit and the linearly interpolated cross-correlation method, respectively.
}\label{fig:lag}
\end{figure}

\begin{figure*}
\centering
\includegraphics[width=0.9\textwidth]{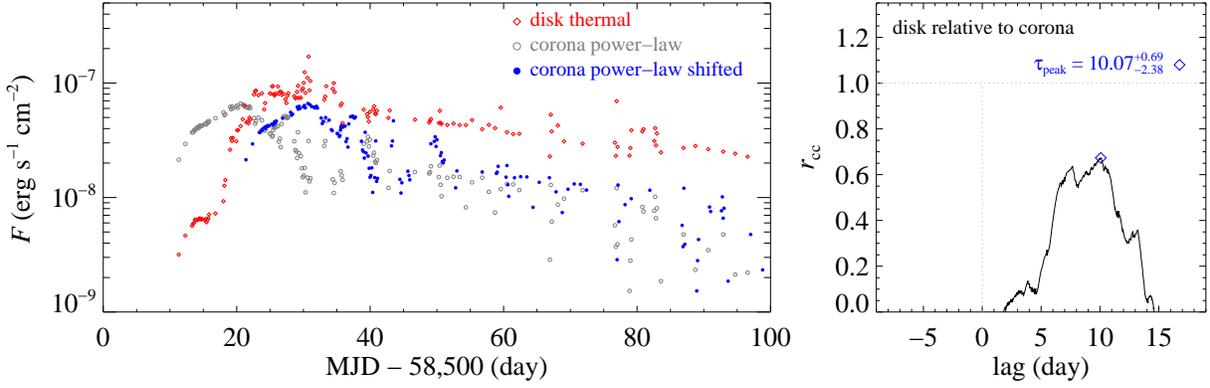}
\caption{The left panel shows the evolving contributions of the disk thermal emission (red diamonds) and the corona power-law emission (gray open circles), while the cross-correlation function for the disk emission with a lag of $\sim 10$ days relative to the corona emission is illustrated in the right panel. To intuitively compare with the disk emission, the corona emission is shifted rightward (blue solid circles in the left panel) by the corresponding lag. \label{fig:flux}}
\end{figure*}

\begin{figure}
\centering
\includegraphics[width=0.5\textwidth]{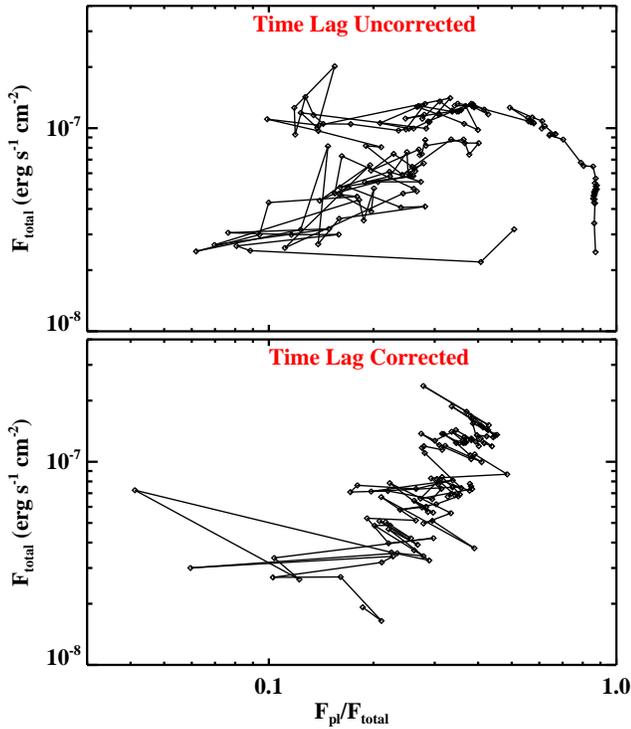}
\caption{Hardness-Intensity Diagram (HID). The hardness here is defined as the ratio of the power-law flux, $F_{\rm pl}$, to the total flux, $F_{\rm total}$($= F_{\rm disk}+F_{\rm pl}$), in 1--10 keV. The ``q''-shaped pattern is displayed in the original HID (top panel), which becomes a linear correlation after having corrected for the time-lag effect (bottom panel).
\label{fig:hid}}
\end{figure}

\begin{figure}
\centering
\includegraphics[width=0.5\textwidth]{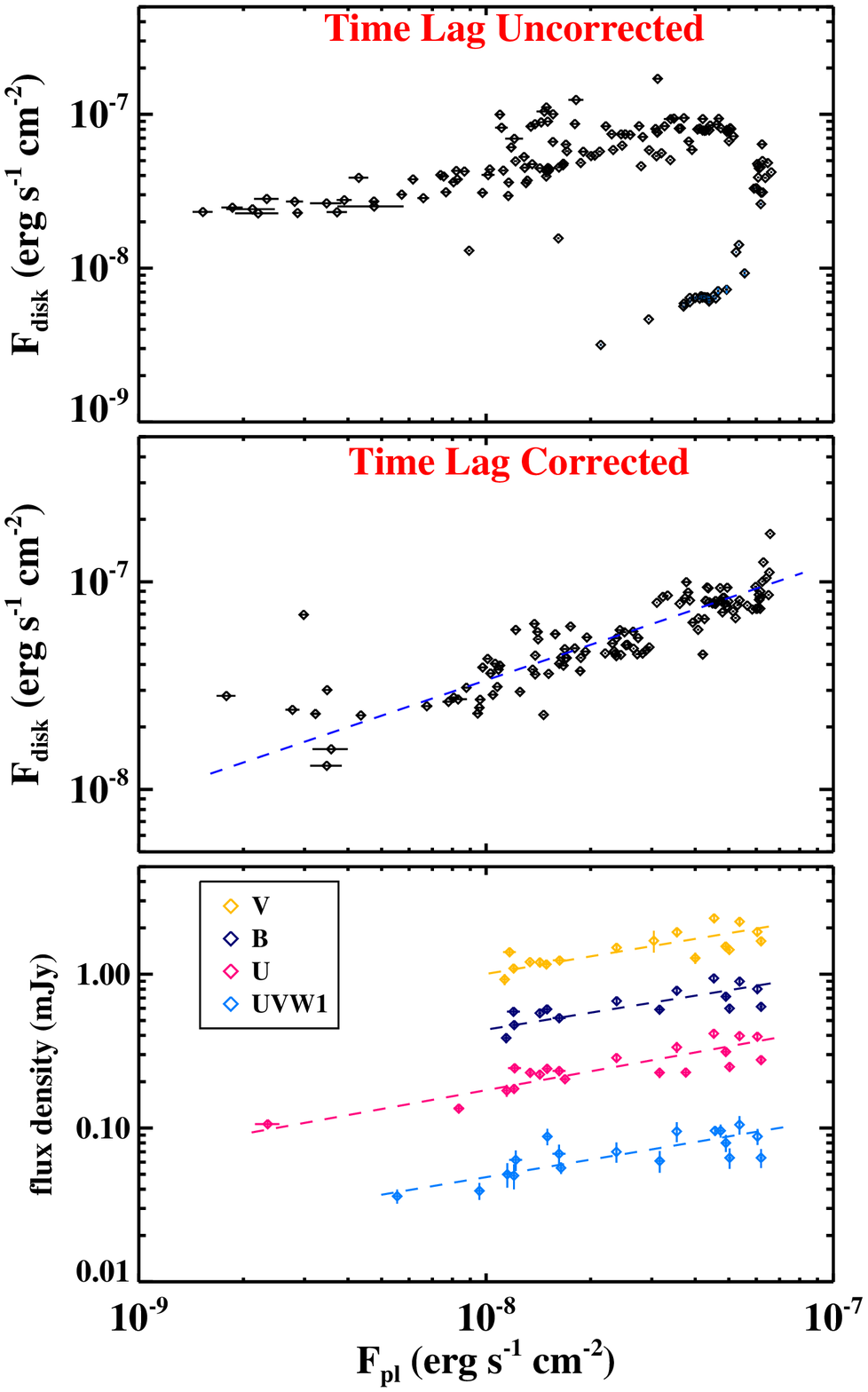}
\caption{Scaling relations. Top and middle panels: correlations between the disk flux and nonthermal flux (in 1--10 keV) without and with the time-lag corrected, respectively. Bottom panel: correlations between the optical flux density derived from \swift{}/UVOT data and the nonthermal flux. All correlations are fit with a power-law function (dashed lines in the middle and bottom panels).
\label{fig:corr}}
\end{figure}

\subsection{{\it Swift} light curves}
The Neil Gehrel \swift{} Observatory is a rapid response multi-wavelength satellite with three main instruments on board: the Burst Alert Telescope, the X-ray Telescope (XRT), and the UV/Optical Telescope \citep[UVOT; ][]{gehrels2004}. It carried out 34 pointings to monitor the primary outburst of \maxi{}. We process both the XRT and UVOT data with the packages and tools available in \textsc{heasoft} version 6.27. Around the peak of the outburst, the observed XRT count rates are more than 1000 cts/s, which are much higher than the normal count rate of the instrument ($\sim 100$ cts/s for the window-timing mode\protect\footnote{\url{https://www.swift.ac.uk/analysis/xrt/xrtpileup.php}}). Technically, a large bright center region should be excluded to handle the significant pile-up effect, and there remains only a small number of photons. Thus, the XRT data are not suitable for the following energy-dependent time-lag estimation.

In this work, we only use the UVOT data that were taken in the image mode. For each observation, we stack the images when there is more than one exposure in order to increase photon statistics. The counts are converted to the flux densities with source aperture radius of 5 arcseconds in the stacked images using the task \textsc{uvotsource}, and a larger neighboring source free sky region is adopted for estimating the background. The Galactic extinction in the direction of \maxi{} is heavy \citep[$A_{\rm V} =12.4$ and $E(B-V) = 4.4$; ][]{schlafly2011}, while the neutral column density inferred from its X-ray spectrum also indicates a moderate extinction for the source \citep[$A_{\rm V} =2.4$; ][]{russell2019a}. Consequently, most of near-ultraviolet emissions are absorbed, resulting in only 17/15/22/20/0/5 detections for the V/B/U/UW1/UM2/UW2 bands, respectively. Therefore, we only consider the V/B/U/UW1 flux densities\footnote{Throughout this Letter, we use the term ``optical emission'' to include also the UW1 flux.} and their relationship to the \hxmt{} data.

\subsection{\hxmt{} data analysis}
\hxmt{}, launched on 2017 June 15, is the first X-ray astronomy satellite of China. It consists of three scientific instruments: the High Energy X-ray telescope (HE, 20--250 keV), the Medium Energy X-ray telescope (ME, 5--30 keV), and the Low Energy X-ray telescope \citep[LE, 1--15 keV; ][]{zhang2020}. The \hxmt{} ToO observations started on 2019 January 27 \citep{chen2019} and ended on 2019 April 27. All data are reduced with the \textsc{hxmtdas} v2.01, and the initial event cleaning is performed with the standard quality cuts: (1) pointing offset angle $< 0.04\degr$; (2) elevation angle $> 10\degr$; (3) value of the geomagnetic cutoff rigidity $> 6$; (4) the time after and to next SAA $> 300$ s. The background is estimated using the tasks LEBKGMAP, MEBKGMAP, and HEBKGMAP in the \textsc{hxmtdas} software. Thanks to the large effective area, \hxmt{} can accumulate enough photons without the pile-up effect in an individual observation. Nine LE light curves are constructed from 1 keV to 10 keV in bins of 1 keV, and count rates are averaged for each individual observation. Meanwhile, the light curves in 10--20 keV and 20--150 keV are extracted from the ME and HE data, respectively (Figure~\ref{fig:lc}).

Broadband coverage of \hxmt{} is helpful in resolving different spectral components. The spectra of \hxmt{} throughout the whole outburst can be fit by an absorbed disk blackbody plus a cutoff power-law model, i.e. {\it tbabs*(diskbb+cutoffpl)} in \textsc{xspec} \citep{arnaud1996}. More detailed spectral investigation will be presented in Zhang~Wei et al. (2021, submitted to ApJ). Figure~\ref{fig:spec} shows an example spectrum in the soft state on 2019 March 8, with the best-fit parameters and errors in 90\% confidence level of $kT = 0.643_{-0.003}^{+0.003}$ keV, $\Gamma = 2.28_{-0.04}^{+0.04}$, and $E_{\rm cut} = 210_{-30}^{+50}$ keV. Our \hxmt{} spectral fitting results are similar to those obtained with {\it NICER} \citep{zhangliang2020}. The unabsorbed thermal and cutoff power-law fluxes are calculated in 1--10 keV with the convolution model {\it cflux}.

\section{Results}

\subsection{Energy-dependent time-lag estimation}
The X-ray properties of \maxi{} are consistent with those of the classical BHXRBs \citep{tominaga2020, jana2020, belloni2020, zhangliang2020}. Visually, multi-band light curves of the outburst exhibit a ``fast-rise slow-decay'' profile, and the soft X-ray flux reached the maximum several days after the hard X-ray one, as shown in Figure~\ref{fig:lc}. Since the \hxmt{} ToO observations have much better sampling than the \swift{}/UVOT data, we quantitatively estimate the time-lag of all light curves relative to the \hxmt{}/HE light curve in two ways. First, in order to avoid possible anomalous fluctuations, we employ a fourth-order polynomial fit to the peak four days before and after the maximum (the gray solid vertical lines in Figure~\ref{fig:lc}). Note that the estimated peak epochs only varies slightly, $< 1$ days, if a third- or fifth-order polynomial is used.

Second, we also estimate the lags of these light curves relative to the HE light curve using the linearly interpolated cross-correlation method \citep[e.g.,][]{peterson1998, sun2018, cai2020}. Figure~\ref{fig:ccf} shows the corresponding cross-correlation functions with the peak lag, $\tau_{\rm peak}$, labled as blue diamonds. The 10-90th percentile uncertainties of $\tau_{\rm peak}$ are inferred from simulated light curves using the flux randomization/random subset selection method, wherein $10^3$ realizations of both light curves with flux measurements are adjusted by random Gaussian fluctuations scaled to the measurement uncertainties. Note that in applying a random subset selection for a light curve, a part of the initial epochs, $\sim 36\%$, would be omitted. Since the \swift{} light curves are very sparse with only $\sim 15$ epochs, the random subset selection would not be performed when estimating the uncertainties of their lags relative to the HE light curve.

Relative to the HE light curve, the peak lag, $\tau_{\rm peak}$, as a function of energy is illustrated in Figure~\ref{fig:lag}, where $\tau_{\rm fit}$ inferred from the polynomial fit is also shown for comparison. Bi-modality of the lag-energy relation is obvious: hard X-rays above around 6 keV and optical emission are synchronized with negligible lags, but soft X-rays below about 6 keV lag behind by about 8-12 days. These results indicate that the soft component does not change in lockstep with the hard X-ray emission. The time-lag, $\sim 10.07_{+0.69}^{-2.38}$ days, between the thermal and the power-law emissions is measured with the linearly interpolated cross-correlation method as well (Figure \ref{fig:flux}). This value is consistent with those directly inferred from the multi-wavelength light curves illustrated in Figure~\ref{fig:lag}.

\subsection{Scaling relations}

In order to compare the scaling relations found in other BHXRBs, we calculate the unabsorbed flux densities in the energy range of 1--10 keV, which is similar to those used in literature \citep[e.g.,][]{russell2006, rykoff2007}. The ``q''-shaped HID pattern is reproduced in the top panel of Figure~\ref{fig:hid}, where the hardness is defined as the ratio of the nonthermal power-law flux, $F_{\rm pl}$, to the total flux, $F_{\rm total}$ \citep[e.g.,][]{dunn2010}. However, when the time-lag effect has been corrected by shifting the disk flux $F_{\rm disk}$ backward (or power-law flux $F_{\rm pl}$ forward) in time by $\sim$10 days and then recalculating the total flux, the hysteresis behavior is eliminated from the ``q''-shaped HID (Figure~\ref{fig:hid}), and the thermal flux monotonically increases with the power-law flux, as shown in the middle panel of Figure~\ref{fig:corr}. The correlation is fit with the power-law function with {\it bces} \citep{akritas1996}: ${\rm log}(F_{\rm disk}) = C+n \times {\rm log}(F_{\rm pl})$, where $C$ is a constant and parameter $n$ is the slope of the correlation. In this way, both 1 $\sigma$ errors of $F_{\rm disk}$ and $F_{\rm pl}$ are considered in the fitting, and $n = 0.57\pm0.03$ is derived.

We fit the optical and the power-law flux correlation with the same methodology, and obtain the slopes $n = 0.38\pm0.06$, $0.37\pm0.06$, $0.41\pm0.04$, and $0.38\pm0.04$, for the V, B, U, and UW1 bands, respectively. The values of these parameters are significantly smaller than the value ($\sim 0.7$) given in the jet scenario \citep{gallo2003, heinz2003}, and are consistent with (or slightly smaller than) the value of 0.5 predicted by the reprocessing theory \citep{van1994, russell2006}.

\section{Discussion and Conclusion}
The 2019 January outburst of \maxi{}, in particular its rise stage, was well traced by \hxmt{} and \swift{} covering visible, UV, soft X-ray, and hard X-ray bands. In this work, we investigate the broadband monitoring data of \maxi{} and have obtained the main results as follows: bi-modality of lag-energy relation (Figure \ref{fig:lag}); elimination of the hysteresis behavior from the HID by taking the time-lag effect into account (Figure \ref{fig:hid}); and the linear correlation between the power-law flux and the optical emission (Figure \ref{fig:corr}).

The optical emissions from BHXRBs in their low-luminosity state can have contributions by the companion star, the jet, or viscously heated disk, resulting in different correlations between the X-ray and optical emissions \citep{russell2006, weng2015}. In contrast, the reprocessing of X-rays is dominant at optical wavelengths in the high-luminosity state \citep[e.g.,][]{van1994}. \cite{rykoff2007} revealed the strong relationship between the optical and the X-ray radiations during the 2006 outburst of XTE~J1817-330, and provided strong evidence to the irradiation scenario. They further pointed out that the optical emission closely tracked the power-law flux and did not track the disk flux. However, the fast-rise of the outburst was missed, and the \swift{} data only covered the  monotonic decline of its 2006 outburst. For \maxi{}, because the rise stage of its 2019 January outburst was caught by both \swift{} and \hxmt{}, we can firmly rule out the direct correlation between the visible emission and the disk flux because of the obvious time-lag (Figure \ref{fig:lag}). Meanwhile, the optical emission is nearly synchronized with the power-law component, and the correlation slopes between them are in the range of $0.37-0.41$ (Figure~\ref{fig:corr}). These values are slightly smaller than 0.5 predicted by the reprocessing theory \citep{van1994, russell2006}, and the deviation could be due to a more complicated geometry of corona in \maxi{}. Alternatively, the synchrotron radiation of nonthermal electrons in the corona sometimes might give non-negligible optical emission, which is predicted to anti-correlate with X-rays \citep{veledina2011, lopez2020}, thus probably causing the slope slightly smaller than 0.5 as we measured.

The time-lags estimated by both the multi-wavelength light curves and the resolved disk plus corona components suggest that the thermal component lags behind the nonthermal component by $\sim 8-12$ days. This is consistent with the viscous timescale for the matter traveling from the outer disk region to the vicinity of the central black hole. Intriguingly, after correcting for the time-lag effect, the ``q'' pattern HID is changed to a tight linear correlation (Figure \ref{fig:hid}). Therefore, we conclude that the observed time-lag between radiations of the accretion disk and the corona leads naturally to the hysteresis effect and the ``q''-diagram observed in \maxi{}. However, we do not know whether this mechanism universally works for other BHXRBs. Until recently, for only a small number of BHXRB outbursts, their fast-rise stages had been caught with high cadence pointed observations but over only limited energy/wavelength bands \citep{koljonen2016, kara2019}. Additionally, it would be hard to estimate the time-lag due to the complicated evolution patterns of thermal/nonthermal components. Finally, we propose a panorama of accretion disk/corona for \maxi{}: the hard X-ray from the corona heats and induces the corresponding optical brightening in the outer disk; thereafter, the enhanced accretion in the outer disk propagates inward, at viscous timescale of $\sim$8-12 days, to the inner disk region where the soft X-rays are produced. Unfortunately, because the very beginning of the outburst was not captured by any telescope, the outburst triggering process is still unknown.

\begin{acknowledgements}
This work made use of the data from the {\it Insight}-HXMT mission and public data from the {\it Swift} data archive. {\it Insight}-HXMT is a project funded by China National Space Administration (CNSA) and the Chinese Academy of Sciences (CAS). The authors acknowledge support from the National Natural Science Foundation of China under grants U2038103, 11873045, 11733009, U1838202, U1938101, U1838201, U1838115, U1838108 and U1938107. SSW acknowledges the financial support by the Jiangsu Qing Lan Project.
\end{acknowledgements}

\software{
\textsc{heasoft},
\textsc{hxmtdas},
\textsc{pyccf},
\textsc{idl}
}

\bibliographystyle{aasjournal}
\bibliography{maxij1348}

\end{document}